\documentstyle[12pt]{article}
\newcommand{\be}{\begin{equation}}
\newcommand{\ee}{\end{equation}}
\newcommand{\bi}[1]{\vspace{-3mm}  \bibitem{#1}}

\begin{document}

{\flushright{ \ IC - 94 - 192 }}
\vskip 4 mm

\centerline{\large International Atomic Energy Agency  }

\vskip 2 mm

\centerline{\large United Nations Educational, Scientific and
Cultural Organization }

\vskip 3 mm

\centerline{\Large INTERNATIONAL CENTRE FOR THEORETICAL PHYSICS}
\vskip 15 mm

\centerline{\Large \bf Dissipative Quantum Mechanics: }
\vskip 3 mm

\centerline{\Large \bf The Generalization
of the Canonical Quantization }
\vskip 3 mm

\centerline{\Large  \bf and von Neumann Equation.}
\vskip 15 mm

\centerline{\Large Vasily E. Tarasov$^{*}$ }
\vskip 8 mm

\centerline{ \it International Centre for Theoretical Physics, }
\vskip 2 mm
\centerline{ \it P.O. Box 586, 34100 Trieste, Italy }
\vskip 8 mm

{ {\bf Summary.} --
Sedov variational principle, which is the generalization of
the least action principle for the dissipative processes is used
to generalize the canonical quantization and von Neumann equation
for dissipative systems (particles and strings).

\vskip 2 mm


\vskip 2 mm

 }

\vskip 10 mm

\centerline{MIRAMARE-TRIESTE}
\vskip 4 mm
\centerline{July, 1994 }

-----------------------------------------

\centerline{ * Permanent address: {\it Theoretical High Energy
Physics Department, Nuclear Physics Institute,} }
\centerline{  \ \ \ \ \ \ \ {\it  Moscow State University,
119899 Moscow, Russia }}
\centerline{ \ \ \ \ \ E-mail: TARASOV@THEORY.NPI.MSU.SU }

\eject

\section{ Introduction.}

In the last two decades great numbers of papers have deal with the
quantum description of the dissipative systems (\cite{Prig1} - \cite{Tarpl})
-- a subject pioneered by Cardirola \cite{Card}.

The dissipative models in string theory are expected to have more broad
range of application. It is caused by the following possibilities:

1) Noncritical strings are dissipative systems in the "coupling constant"
phase space \cite{EMN,Nanop}. In this case, dissipative forces are defined
by non-vanishing beta-functions of corresponding coupling constant and by
Zamolodchikov metric \cite{EMN,Nanop}.

2) Problems of quantum description of black holes on the two-dimensional
string surface lead to the necessity of generalization of von Neumann
equation for dissipative systems \cite{EHNS,EMN,Nanop}.

3) The motion of a string (particle) in affine-metric curved space
is equivalent to the motion of the string (particle) subjected to
dissipative forces on Riemannian manifold \cite{Tarpl,Tarmpl}.
So the consistent theory of the string in the affine-metric curved
space is a quantum dissipative theory \cite{Tarpl,Tartmf2,Taryf,Tarmpl}.

But the quantum descriptions of the dissipative
systems (particles, strings,...) have well known ambiguities
\cite{Lem,Hav1,Hav2,Edv,Hojm,Mess,Ber1,Ber2,Kor1,Kor2}.
Let us describe briefly some approaches to the quantum description of
dissipative systems. I hope you'll forgive me the subjectivity
in the description of the published works.

{\bf 1.1. Quantum kinetics.}

Vectorial Newtonian mechanics describe the motion of mechanical systems
subjected to forces. The forces usually are divided into potential and
dissipative forces. The Newtonian approach does not restrict the nature of
the force \cite{Lanc}. Variational Lagrangian and Hamiltonian mechanics
describe the systems subjected to the potential forces only \cite{Lanc}.
Dissipative forces are beyond the sphere of the variational principles of
least action \cite{Sed1}. For this reason the statistical mechanics cannot
describe the irreversible processes. It is caused
by the absence of Liapunov function in the phase
space in Hamiltonian mechanics (Poincare-Misra theorem \cite{Misr,Prig1}).
To describe the dissipative and irreversible processes we must introduce an
additional postulate in statistical mechanics (for example, the Bogolubov
principle of weakening correlation \cite{Bog1} and the hypothesis of the
relaxation time hierarchy \cite{Bog2} ). Therefore
these processes are considered within framework of the physical kinetics
\cite{Akh,Lib}.

It is known that the initial point of the quantum mechanics formalism is
Hamiltonian mechanics \cite{Dir}. Therefore the quantum theory describes
physical objects in the potential force fields only. The
irreversible and dissipative quantum dynamics outside the framework of
quantum mechanics. The quantum description of the dissipative and
irreversible processes is the aim of quantum kinetics which is
the quantum statistics with the additional physical postulates.

Quantum description of dissipative systems within framework of quantum
kinetics is very popular and successful, but it is not valid in the
fundamental theories such as string theory.

{\bf 1.2. Generalization of von Neumann equation.}

The other way to attack
the problem of quantum dynamical description of the dissipative
systems is the generalization of quantum statistics
\cite{Ber1,Ber2,EHNS,Kor1,Kor2,Mess,Lind,Gor1,Gor2}. An important
property of the dissipative and irreversible processes is the
increase of the entropy during these processes. Nevertheless, the
quantum mechanical evolution equation for statistical
("density-matrix") operator, the quantum Liouville (called von
Neumann) equation keeps the entropy invariant \cite{Neum}.

The generalized von
Neumann equations proposed until now, which should describe dissipative
and irreversible processes are derived by the addition the
superoperator which acts on the statistical operator and describes
dissipative part of time evolution. The linear generalized von
Neumann equations are connected with master (Pauli) equation
of the quantum kinetics \cite{Akh} and with the quantum dynamical semigroup
\cite{Lind,Gor1,Gor2}.
Note that the total time derivative of the dissipative system
operator does not satisfy the Leibnitz rule, i.e. it is not the
derivative operator, and called the dissipative operator \cite{Brat1}.

Messer and Baumgartner \cite{Mess}, Beretta, Gytopoulos, Park, Hatsopoulos
\cite{Ber1,Ber2} and Korsch, Steffen, Hensel \cite{Kor1,Kor2} proposed the
nonlinear generalizations of von Neumann equation corresponding to the
nonlinear Schroedinger equations introduced by Gisin \cite{Gis},
Kostin \cite{Kost} to describe dissipative systems.

Note that proposed generalizations
of the von Neumann equation are derived heuristically. For a given set
of generalized equations different requirements for superoperator
exist. Most of the requirements proposed until now, which should
determine superoperator uniquely are not unique themselves and so one
has to deal with the problems arising from these ambiguities.
The superoperator form is not determine uniquely.
Moreover, the generalizations of the von Neumann equation are not
connected with classical Liouville equation for dissipative systems
\cite{Liuv,Steb,Prig2,Fron}. Therefore the quantum description of
the dissipative system dynamics used the generalizations of von Neumann
equation proposed until now is  ambiguous.

{\bf 1.3. Problems of canonical quantization.}

Let us mention some papers showed that canonical quantization of the
dissipative systems is impossible.

{\bf 1.3.1.}
Lemos \cite{Lem} proved that canonical quantization is impossible in
the presence of dissipative forces. So dissipative systems are not compatible
with canonical quantization commutation relations. Note
that Lemos considered the total time derivatives of the commutation
relations for the coordinates and momentums, used the Jacobi
identity and the dissipative equations of motion of Heisenberg
operator.

It is easy to see that quantum equation of motion for dissipative
system is not compatible with Heisenberg algebra. Let us consider
the quantum Langevin equation \cite{Haken}
\[ \dot a = (- \imath \alpha - \beta ) a + f(t) \]
\[ \dot \dagger a = (- \imath \alpha - \beta ) \dagger a + \dagger f(t) \]
We have
\[ [\dot a, \dagger a] + [a, \dot \dagger a] \ = - 2 \beta \]
On the other hand, the total time derivative of the Heisenberg algebra
\[ [a, a] \ = \ [\dagger a, \dagger a] \ = \ 0 , \ [ a, \dagger a] = 1 \]
and Leibnitz rule lead to the following
\[ [\dot a, \dagger a] + [a, \dot \dagger a] \ = 0 , \]
i.e. quantum dissipative equations of motion are not compatible with
canonical commutation relations and Heisenberg algebra.

{\bf 1.3.2.}
As is known that the equation of motion is the Euler-Lagrange equation
based on local a Lagrangian function when the Helmholtz conditions are
satisfied \cite{Helm}. Havas \cite{Hav2} considered a
general theory of multipliers which allows (by using the Helmholtz
conditions) a Lagrangian formulation for a broad class of the equation
of motion of the dissipative systems, which cannot
fit into Lagrangian mechanics by usual approach.

Havas therefore noted that the quantization of systems described by
Lagrangian of the above type is either impossible or ambiguous \cite{Hav1}.
This is follow from the fact that in classical mechanics of the
dissipative systems there are many different Lagrangians and
Hamiltonians leading to the same equations of motion  \cite{PhysRep}.
So we  don't know which of the possible Lagrangians is corrected
and one to choose for quantization procedure.

{\bf 1.3.3.}
Edvards \cite{Edv} showed that, although classical Hamiltonians are
necessary for canonical quantization, their existence is not sufficient
for it. The quantization of the Hamiltonian which is not canonically
related to the energy is ambiguous and therefore the results are
conflicting with physical interpretations.

It is not sufficient for the
Hamiltonian to generate the equation of motion, but Hamiltonian must
also be necessarily related via canonical transformation to the total energy
of the system. However, this condition can only be met by
conservative systems, thus excluding dissipative systems from
possible canonical quantization \cite{Edv}.

{\bf 1.3.4.}
Hojman and Shepley \cite{Hojm} started with classical equation of motion
and set very general quantization conditions (relation that the coordinate
operators commute). The total time derivative of this commutation
relation was considered. It was showed that commutator of the coordinate
operator
and the velocity operator form a symmetric tensor operator.
They proved that classical analog of this tensor operator is
a matrix which inverse matrix satisfies the Helmholtz conditions.
Using the Jacobi identity for the coordinate and the velocity, Hojman and
Shepley conclude the following:
the general quantization condition implies that the equation of motion
 is equivalent to Euler-Lagrange equation of the Lagrangian.
It is not sufficient for Lagrangian merely to generate the equation of
motion, but it must necessarily give rise to Hamiltonian which is canonically
related to the physical energy of the system \cite{Edv}.

{\bf 1.4. Nonassociative Lie-admissible quantization.}

The generalization of the canonical quantization of the dissipative systems
was proposed by Santilli \cite{Sant1,Sant2,Sant3}. Santilli
showed that the time evolution law of dissipative equation
not only violates Lie algebra law but actually does not characterize an
algebra. Therefore Santilli suggested,  as a necessary condition
to preserve the algebraic structure, that the  quantum dynamics of the
dissipative systems should be constructed within the framework of the
nonassociative algebras. This is exactly the case of the noncanonical
quantization at the level of the nonassociative Lie-admissible
(or Lie-isotopic) enveloping algebra worked out by Santtili, as well
as flexible case suggested by Okubo \cite{Okub}. The Lie-admissible
generalization of the "density-matrix" operator and von Neumann
equation was considered by Mignani \cite{Mig}.

The quantization of the dissipative systems was proposed by Santilli
\cite{Sant1,Sant2,Sant3} as an operator image of the
Hamiltonian-admissible and Birkhoffian generalization of the
classical Hamiltonian mechanics. The generalized variations used by
Santilli \cite{Sant4} to consider the dissipative processes in the field of
the holonomic variational principles are connected with the generalized
multipliers suggested by Havas \cite{Hav2} and therefore lead to
an ambiguity in generalized variations.

{\bf 1.6. Nonholonomic variational principle. }

Sedov \cite{Sed1}- \cite{Sdch1} suggested
the variational principle which is the generalization of the least
action principle for the dissipative and irreversible processes. The
holonomic and nonholonomic functionals are used to include the dissipative
processes in the field of the variational principle. The suggested
variational principle was used by Sedov, Chipkin  \cite{Sdch2},
Berdichevskii  \cite{Berd}, Jelnorovich  \cite{Jeln,Sdjl}  to
construct the phenomenological models of the continuous media with
irreversible processes.

On of the approaches used a path function with
the properties of an action in order to describe quantum systems with
friction was suggested by Battezzati in the paper \cite{Batt}.

{\bf 1.5. We can conclude the following:}

{\it
1. Canonical quantization of the dissipative systems is impossible if
all operators in quantum theory are associative.

2. In the consistent dissipative quantum theory equation of motion must be
compatible with canonical commutation relations.

3. Hamiltonian must be canonically related to the physical energy of the
dissipative system.

4. Total time derivative of the dissipative system operator does not satisfy
the Leibnitz rule.

5. Generalization of the von Neumann equation must be connected with classical
Liouville equation for dissipative systems.

6. Dissipative systems can be described within framework of
the nonholonomic variational principle.}

In this paper we consider some main points of the quantum description
of the dissipative systems which take into account these conclusions.

{\bf 1.6. Nonholonomic variational principle and quantum description of the
dissipative systems. }

Nonholonomic principle was suggested in \cite{Tar1,Taryf,Tartmf1} to
generalize the classical mechanics in phase space.
The classical mechanics of the dissipative systems in the phase space
suggested in this paper can be used to consider the generalizations of
canonical quantization for dissipative systems and von Neumann equation.

In order to solve the problems of the quantum description of dissipative
systems we suggest to introduce an operator $W$ in addition to usual
(associative) operators.
{\it The suggested operator algebra does not violate Heisenberg algebra
because we extend the canonical commutation relations by introducing
an operator of the nonholonomic quantity in addition to the usual
(associative) operators of usual (holonomic) coordinate-momentum functions.
That is the coordinate and momentum satisfy the canonical commutation
relations.} To satisfy the generalized commutation relations the operator
$W$ of nonholonomic quantity must be nonassociative nonlieble (does not
satisfied the Jacobi identity) and Lie-nonaddmisible  operator
\cite{Taryf,Tartmf2}. As the result of these properties the total
time derivative of the multiplication and commutator of the operators
does not satisfies the Leibnitz rule.
This lead to compatibility of quantum equations of motion for dissipative
systems and canonical commutation relations.
The suggested generalization of the von Neumann equation is connected
with classical Liouville equation for dissipative systems.

The dissipative quantum scheme suggested in  \cite{Taryf,Tartmf2} and
considered in this paper allows to formulate the approach to the quantum
dissipative field theory and quantization of the phenomenological models
of continuous media. As an example of the dissipative quantum field theory
the sigma-model approach to the quantum string theory was considered in the
recent papers \cite{Tarpl,Tarmpl}.
Conformal anomaly of the energy momentum tensor trace for closed
bosonic string on the affine-metric manifold and two-loop metric
beta-function for two-dimensional nonlinear dissipative sigma-model
was calculated \cite{Taryf,Tarpl,Tarmpl}.

\section{ Nonholonomic Variational Principle.}

{\bf 2.1.}
The equations of motions of the mechanical systems in n-dinensional
configurational space are
\be
\label{1}
D_i T(q,u,t)  - f_i = 0
\ee
where T is the kinetic energy, which can be written in the form
\be
T(q,u,t) = \frac{1}{2} {a_i}_j (q,t) u^i u^j + a_i (q,t) u^i +
a_0 (q,t)
\ee
\be
D_i \equiv \frac{d}{dt} \frac{ \partial}{ \partial u^i } - \frac{
\partial}{ \partial q^i}
\ee
$ i;j = 1,...,n ; u^i \equiv { {dq^i}/{dt}} $ and $ f_i = f_i (q,u,t) $
is the sum of external forces. In general case, $ f_i $ is the sum
of the potential $ f^p_i $ and the dissipative $ f^d_i $ forces.
The potential force is the force for which a function $ V = V (q,u,t) $:
\be
\label{2}
D_i V = \, f^p_i
\ee
exists. The dissipative force $ f^d_i $ is the force which cannot be
written in the form (\ref{2}). Then the Euler-Lagrange equations take the form
\be
D_i L - f^d_i = 0
\ee
where $ L = L(q,u.t) \equiv T-V $ is Lagrangian. In the dissipative case
( $ f^d \not\equiv 0 $ ),
the equation (\ref{1}) cannot be followed from the least action
principle \cite{Lanc}:
\be
\label{act}
\delta S(q) \equiv \delta \int dt \ L(q,u,t) = 0
\ee

{\bf 2.2.} Helmholtz conditions for the differential equations of motion
\be
  \label{eq}
  F_k(t,q,u,...,q^{(m)}) \ = \ 0
\ee
where $q^{(m)} = d^m q /  dt^m \ $, are equivalent \cite{Berdich}
to a holonomic condition
for the functional $S(q)$, defined by the variational equation
\be
 \delta S(q) = \int^{t_2}_{t_1} \ dt \   F_k(t,q,u,...,q^{(m)}) \ \delta q^k
\ee
This functional is a holonomic functional if and only if
\be
\oint_{\Gamma} \delta S(q) = \int_{\Sigma} (\delta \delta^{/} S(q) -
\delta^{/} \delta S(q)) \ = \ 0
\ee
The holonomic condition is given by \cite{Berdich}
\be
\label{hol}
\delta \delta^{/} S(q) - \delta^{/} \delta S(q) \  = \ 0
\ee
It is easy to see that this condition lead to Helmholtz conditions
\be
\label{Helm1}
\frac{\partial F_k}{\partial q^l} \ - \ \frac{\partial
F_l}{\partial q^k} \ - \ \sum^{m}_{s=1} [-1]^s \frac{d^s}{dt^s}
\frac{\partial F_l}{\partial q^{k(s)}} \ = \ 0
\ee
\be
\label{Helm2}
\frac{\partial F_k}{\partial q^{l(i)}} \
- \ \sum^{m}_{s=i} [-1]^s \frac{d^s}{dt^s} (^s_i)
 \frac{\partial F_l}{\partial q^{k(s)}} \ = \ 0
\ee
If the conditions (\ref{hol}) or (\ref{Helm1},\ref{Helm2}) are satisfied
then a local Lagrangian function exists and the integral of this function is
the holonomic functional called action
\be
S(q) = \int^{t_2}_{t_1} \ L (t,q,u,...,q^{(m)}) \ dt
\ee
In this case the equations (\ref{eq}) can be derived from the least action
principle (\ref{act}). Note that, if the Lagrangian exists we can define the
metric in the $(n+1)$-dimensional configurational space \cite{Lanc}. So the
motion on the metric manifold is equivalent to the motion of the non
dissipative system \cite{Lanc,Tarpl}

{\bf 2.3.} The basic variational principle for dissipative processes is the
nonholonomic principle suggested by Sedov \cite{Sed1}-\cite{Sdch1}.
It is a generalization of the least action principle.
The Sedov variational principle has the form:
\be
\delta S(q) + \delta  \tilde W (q) = 0
\ee
where S(q) is the holonomic functional called action and $ \tilde W(q) $
is the nonholonomic functional. The nonholonomic functional is
defined by the nonholonomic equation. Let a variation of the
nonholonomic functional be linear in the variations
$ \delta q^i $ and $ \delta u^i $ that is
\be
\delta \tilde W = \delta \int dt \ w(q,u) = \int dt \ (f_i (q,u) \delta
q^i + g_i (q,u) \delta u^i )
\ee
where $ f_i $ and $ g_i $ are the vector functions in the
configurational space. Let us consider the Hamiltonian approach
to the variational classical mechanics with dissipative forces.

\section{ Dissipative Mechanics in Phase Space.}

{\bf 3.1.} One direct corollary of the nonholonomic variational
principle is the
following dissipative equation of motion
\be
\frac{d}{dt}( \frac{ \partial L(q,u)}{\partial u^i}+ g_i (q,u))
= \frac{ \partial L(q,u)}{ \partial q^i} + f_i (q,u) \qquad
\frac{ dq^i}{dt} = u^i
\ee
Let us define a canonically conjugate momentum by the equations
\be
p_i \equiv \frac{ \partial L(q,u)}{\partial u^i} + g_i (q,u)
\ee
and represent this relation in the form $ u^i = v^i (q,p) $.
The Hamiltonian is given by
\be
h(q,p) = p_i v^i (q,p)- L(q,v(q,p))
\ee
If we consider the variation of the Hamiltonian, we obtain the
dissipative Hamiltonian equations of motion
\be
\label{3}
\frac{dq^i}{dt} = \frac{\delta(h-w)}{\delta p_i} \quad ; \qquad
\frac{dp_i}{dt} = - \frac{\delta(h-w) }{\delta q^i}
\ee
where
\be
 \delta w(q,p)= \delta w(q,v(q,p))= w^q_i \delta q^i + w_p^i \delta p_i
\ee

{\bf 3.2.}  Let the coordinates $ z^k , \ ( k =1,..., 2n), \ $ where
$ \ z^i=q^i, \ z^{n+i}=p_i \ \ (i=1,...,n) \ $ and $ \ w, t $ of the
(2n+2)-dimensional extended phase space be connected by the equations
\be
\label{4}
d w \ - \ a_k (z,t) \ d z^k \  = \ 0
\ee
where $ a_k \ ( k=1,...,2n) $ are the vector functions in phase space.
Let us call the dependence $w$ on the coordinate $q$ and momentum $p$ the
holonomic-nonholonomic function and denote $ w=w(z) \in \Phi $.
If the vector functions satisfy the relation
\be
\label{5}
\frac{\partial a_k (z) }{\partial z^l} = \frac{\partial a^l
(z)}{\partial z^k }
\ee
where k;l= 1,...,2n , the coordinate $w$ is the holonomic function ($ w
\in F $). By definition, if these vector functions don't satisfy the
relation (\ref{5}) the object $w(z)$ we call the nonholonomic function
($ w \in \tilde F $). Let us define the generalized Poisson brackets
for $ \forall a,b \in \Phi $ in the form:
\be
[f,g] \equiv \frac{\delta f }{\delta q^i}\frac{\delta
g }{\delta p_i} - \frac{\delta f }{\delta p_i} \frac{\delta g}{\delta q^i}
\ee
The basic properties of the generalized Poisson brackets:

$ \ $

$ 1) Skew-symmetry:  \quad \forall f,g \in \Phi \qquad
[f,g] = - [g,f] \in F ; $

$ \  $

$ 2) Jacobi \ identity:  \quad \forall f, g, s \in F \qquad
J[f,g,s] = 0; $

$ \  $

$ 3) Nonliebility:  \quad \forall f,g,s, \in \Phi : \ f \vee g \vee s
\in \tilde F \qquad  J[f,g,s] \not \equiv 0; $

$ \ $

$ 4) Leibnitz \ rule:  \quad \forall f,g \in \Phi \quad
\frac{ \partial }{ \partial t } [f, g] = [ \frac{ \partial }{ \partial t }
  f, g] + [f ,  \frac{ \partial }{ \partial t }  g]; $

$ \ $

$ 5) Distributive \ rule:  \forall f,g,s \in \Phi \quad
[ \alpha f + \beta g ,s ] = \alpha [f,s] + \beta [ g ,s ] $

$ \ $

where

\[ J[f,g,s] \equiv [f,[g,s]] + [g,[s,f]] + [s,[f,g]]  , \]
$\alpha \ $  and $ \ \beta $ are the real numbers. It is
easy to verify that this properties of the generalized Poisson brackets
for the holonomic
functions coincide with properties of the usual Poisson brackets
\cite{Lanc}. Let us consider now the characteristic properties of the
physical quantities:

$ \ $

$ 1) \quad [p_i ,p_j ] = [q^i ,q^j ] = 0 \qquad and
\qquad [q^i ,p_j ] = \delta^i_j $

$ \ $

$ 2) \quad [w, p_i ] = w^q_i \qquad and
  \qquad [w, q^i ] = - w_p^i  \qquad i \not \equiv j , \qquad [w, w] = 0 $

$ \ $

$ 3) \quad [[w,p_i ],p_j ] \not \equiv [[w,p_j ], p_i ]
\quad or \quad  J[p_i ,w,p_j ] = \omega_{ij} \not \equiv 0
 \qquad i \not \equiv j $

$ \ $

$ 4) \quad [[w, q^i ], q^j ] \not \equiv [[w, q^j ], q^i ]
\quad or \quad  J[q^i ,w,q^j ] = \omega^{ij} \not \equiv 0
 \qquad i \not \equiv j $

$ \ $

$ 5) \quad [[w, q^i ], p_j ] \not \equiv [[w, p_j ], q^i ]
\quad or \quad  J[q^i ,w,p_j ] = \omega^i_j \not\equiv 0 $

$ \ $

where
\be
\omega^i_j \ \equiv \ \frac{\partial w^q_j}{\partial p_i} -
\frac{ \partial w_p^i }{ \partial q^j } \ =  \
 \frac{\delta^2 w}{\delta p_i \delta q^j} - \frac{\delta^2 w
}{\delta q^j \partial p_i}
\ee
This object $ \omega^{kl} \ (k,l=1,...,2n) \ $ characterizes
deviation from the condition of integrability (\ref{5}) for
the equation (\ref{4}) and by the Stokes theorem
\be
\oint_{\partial M} \delta w =  \int_{M} \omega^{kl} \ dz^k \wedge dz^l
 \not \equiv 0
\ee
Note that $w$ is the nonholonomic object if one of  $\omega^{kl}$ is not
trivial.
Therefore some of the properties 3-5 can be not satisfied but one of
it must be carry out if we consider the dissipative processes.

If we take into account generalized Poisson brackets the equation of
motion in phase space for dissipative systems (\ref{3}) takes the form
\be
\label{6}
\frac{dq^i}{dt} = [q^i,h-w] \qquad \frac{dp_i}{dt} = [p_i, h-w]
\ee
The total time derivative of the physical quantity $A = A(q,p,t) \in F$
is given by
\be
\label{7}
\frac{dA(q,p,t)}{dt} = \frac{\partial A(q,p,t)}{\partial t } + [A, h - w]
\ee
The equation of motion (\ref{6}) can be derived from the
equation (\ref{7}) as a particular case. Note that any term which added to
the Hamiltonian $h$ and nonholonomic object $w$ does not change
the equations of
motions (\ref{6}), (\ref{7}). This ambiguity in the definition of the
Hamiltonian is easy to avoid by the requirement that Hamiltonian must
be canonically related to the physical energy of the system \cite{Edv}
\be
 \label{en}
 [w,q^i] \ = \ 0
 \ee

It is easy to see that total time derivative of the generalized
Poisson brackets does not satisfies the Leibnitz rule
\be
\frac{d}{dt} [f, g] = [ \frac{d}{d t } f, g] + [f , \frac{d}{dt} g]
 + J[f,w,g]
\ee

{\bf 3.3.} Let us consider the solution of the equation (\ref{6}) in the form
\be
 q^i = q^i (q_0,p_0,t) \qquad p_i = p_i (q_0,p_0,t)
\ee
We assume that the points of the volume
\[ V_0 = \int \delta q_0 \delta p_0 \]
in the phase space are  initial points at the moment $t
= t_0$ . Then the equations (\ref{6}) transform the volume $V_0$ to
the volume
\[ V = \int \delta q \delta p = \int I \ \delta q_0 \delta
p_0 ,\]
where
\[ I \ = \ \frac{\partial (q,p)}{\partial (q_0,p_0)} \ = \
\frac{\partial q^i}{\partial q^k}\frac{\partial p_i}{\partial p_k}
 - \frac{\partial p_i }{\partial q^k}\frac{\partial q^i}{\partial p_k}
\  = \ [q^i,p_i]_0 .\]
The following equation is easily verified
\be
\label{8}
\frac{dV}{dt} = \int  \delta q \delta p \ \omega(t,q,p)
\ee
where
\[ \omega (t,q,p) \ = \ \sum_{i=1}^{n} \omega^i_i \ =
\ \sum_{i=1}^{n} \ J[q^i, w, p_i] \]

The fundamental hypothesis of the statistical mechanics \cite{Gibs,Prig1} is
that the state at the moment t is defined by the distribution function
$ \rho (q,p,t) $, called density, which satisfies the normalization condition
\be
\label{9}
\int dq dp \ \rho (q,p,t) = 1
\ee
The average of the physical quantity A(q,p,t) is defined \cite{Prig1} by
\be
<A(t)>_{\rho(t)} = \int dq dp \ \rho (q,p,t) \ A(q,p,t)
\ee
Using for  equation (\ref{9})  formula (\ref{8}), we obtain the dissipative
analogue of the Liouville equation \cite{Liuv,Fron,Prig1}:
\be
\label{10}
\frac{d}{dt}  \rho (q,p,t) \ = \ - \ \omega (t,q,p) \rho (q,p,t)
\ee
or
\be
\imath \frac{\partial}{\partial t}  \rho (q,p,t) \ = \ \hat L \rho (q,p,t)
\ee
where
\be
\hat L = \imath ( \ \frac{\delta (h-w) }{\delta q^k}
\frac{\partial}{\partial p_k} - \frac{\delta (h-w)}{\delta p_k}
\frac{\partial}{\partial q^k} - \omega (t,q,p) \ )
\ee
called Liouville operator \cite{Prig1}. In addition to the Poincare-Misra
theorem \cite{Prig1} can be obtained the statement:
" There exists the Liapunov function of the coordinate and momentum
in the dissipative Hamiltonian mechanics". Let us define the
function $ \eta (q,p,t) \equiv -  ln \rho (q,p,t) $ and assume
$ \omega > 0 $. The equation (\ref{10}) shows that
\[ {d}/{dt} \eta (q,p,t) \ = \ \omega (t,q,p) \]
and the function $\eta$ satisfies the relations $ {d \eta}/{dt} > 0 $.
It is convenient to introduce the entropy of the distribution defined
as follows
\be
s \ \equiv \ < \eta > \ = - \int \delta q \delta p \ \ \rho (q,p,t)
\ ln \rho (q,p,t)
\ee
The relation $ {ds}/{dt} > 0 $ is easily verified. In the general case, any
function $ f(q,p,t) $ which is the composite function $ f(q,p,t) =
g(\rho (q,p,t)) $ and satisfies the relation $  \omega \
(\partial g ( \rho )) / (\partial \rho) < 0  \ (\forall t) $ is the
Liapunov function, that is $ (df)/(dt) > 0 $.
A crucial point is that the condition $ \quad \omega > 0 \ $ or $ \
\omega \ (\partial g ( \rho )) / (\partial \rho) < 0  $ is not
necessary \cite{Steb}.

\section{Quantum Dissipative Mechanics.}

{\bf 4.1.}
In order to solve the problems of the quantum description of dissipative
systems we suggest to introduce an operator W in addition to usual
(associative) operators.
Let us use the usual rule of definition of the quantum physical
quantities which have the classical analogues \cite{Dir}: If we consider
the operators A,B,C of the physical quantities a,b,c  which
satisfy the classical Poisson brackets $ [a,b] = c $, then
the operators must satisfy the relation: $ [A,B] \equiv (AB) - (BA) =
\imath \hbar C $. If we take into account the characteristic
properties the physical quantities operators are defined by the
following relations:

$ \ $

$ 1) \quad [ Q^i, Q^j ] = [ P_i, P_j ] = 0 \qquad [ Q^i, P_j ] =
\imath \hbar \delta^i_j $

$ \ $

$ 2) \quad [ W, P_i ] = \imath \hbar W^q_i \qquad [ W, Q^i ] = -
\imath \hbar W_p^i  \qquad [W, W] = 0 $

$ \ $

$ 3) \quad [[W, P_i], P_j] \not \equiv [[W, P_j], P_i] \quad i \not \equiv j
\quad or \quad J[P_i, W, P_j ] = \Omega_{ij} \not \equiv 0 $

$ \ $

$ 4) \quad [[W, Q^i], Q^j] \not\equiv [[W, Q^j], Q^i] \quad i \not \equiv j
\quad or \quad J[Q^i, W, Q^j ] = \Omega^{ij}  \not \equiv 0 $

$ \ $

$ 5) \quad [Q^i ,[W, P_j ]] \not \equiv [P_j, [W, Q^i]] \quad or \quad
J[Q^i, W, P_j ] = \Omega^i_j \not \equiv 0 $

$ \ $

where
\[ J[A,B,C] = -{1}/(\hbar^2) \ ( \ [ A [B C]] + [B [C A]] + [C
[A B]] \ ) \]
and $ \ Q^{\dagger} = Q ; \ P^{\dagger} = P; \  W^{\dagger} = W \ $.
{\it Let us require that the canonical quantum commutation rules be
a part of this rule.} To satisfy the commutation relations and canonical
commutation rules for the operator of the holonomic function the operators
of the nonholonomic quantities must be nonassociative. It is sufficient
to require that the operator W satisfies the following conditions:

 1) \quad  left \ and \ right \ associativity:

 \[ (Z^k,Z^l,W) =(W,Z^k,Z^l) = 0 \]

 2) \quad left-right \ nonassociativity:

\[ (Z^k,W,Z^l) \not \equiv 0  \quad if \quad k \not \equiv l \]

where $ k;l = 1,..,2n ; \ Z^i=Q^i \ $  and $ \ Z^{n+i}=P_i ; \ i=1,...,n ; $
\[ (A,B,C) \ \equiv \ (A(BC)) - ((AB)C) \]
called associator.

{\bf 4.2.} The state in the quantum dissipative mechanics can be represented
by the "matrix-density" (statistical) operator $ \rho (t) $ which
satisfy the condition $ \rho^{\dagger} (t) = \rho (t) $.
The time variations of the operator of physical quantity $ A(t)
\equiv A(Q,P,t) $ and of the operator of state $ \rho (t) $
are written in the form
\be
\label{11}
\frac{dA}{dt} = \frac{\partial A}{\partial t} +
 \frac{\imath}{\hbar} [ H-W, A ]
\ee
\be
\label{012}
\frac{d \rho }{dt} = - \frac{1}{2} [ \ \rho \ , \ \Omega \ ]_{+}
\ee
\be
\label{12}
\frac{\partial }{\partial t} \rho \  = \
\frac{\imath}{\hbar} [ \rho, H ] \ + \
\frac{\imath}{\hbar} [ W, \rho ] \ -  \
\frac{1}{2} [ \ \rho \ , \ \Omega \ ]_{+}
\ee
where anticommutator $ [ \  , \  ]_{+} $ is the consequence of the
hermiticity for the density operator $\rho $ and for the operator
$ \Omega $, which is defined by
\[ \Omega \ = \ \sum_{i=1}^{n} \Omega^i_i \
= \ \sum_{i=1}^{n} \ J[Q^i, W, P_i]  \]

The solution of the first equation may be written in the form
\be
 A(t) = S(t,t_0) A(t_0) S^{\dagger} (t,t_0)
\quad where \qquad
S(t,t_0) = Texp \frac{\imath}{\hbar} \int^{t}_{t_0}d \tau \ (H-W)(\tau)
\ee
T-exponent is defined as usual, but we must take into account the
following flow chart
\[ exp \ A = 1 + A + \frac{1}{2} (AA) + \frac{1}{6} ((AA)A) + \frac{1}{24}
(((AA)A)A) + ...  \ \ . \]

The solution of the equation (\ref{12}) is given by
\be
\rho (t) = U(t,t_0) \rho (t_0) U^{\dagger} (t,t_0) \quad where \quad
U(t,t_0) = Texp \frac{1}{2} \int^{t}_{t_o} d \tau \ \Omega ( \tau )
\ee
In this way the time evolution of the physical quantity operator is
unitary and the evolution of the state operator is nonunitary.
It is easy to verify that the pure state  at the moment
$ t = t_0 $ ( $ \rho^{2} (t_0) = \rho (t_0) $ )
is not a pure state at the next time moment $ t \not \equiv t_0 $. We can
define the entropy operator $ \eta $ of the state $ \rho (t) $   :
$ \ \eta (t) = - \ ln \rho (t) $. The entropy operator satisfies the
equation
\[ \frac{d}{dt}  \eta (t) \  = \ \Omega \]

It is easy to see that the commutator with nonassociative operator $W$
and the total time derivative of both the quantum Poisson brackets
and of the multiplication of the two operators do not satisfy the
Leibnitz rule
\be
[AB, W] =  A [  B, W] + [A , W ] B \ + \ (A,W,B)
\ee
\be
\frac{d}{dt} [A, B] = [ \frac{d}{d t } A, B] + [A , \frac{d}{dt} B]
\ + \ J[A,W,B]
\ee
\be
\frac{d}{dt} (A B) = (( \frac{d}{d t } A) B) + (A ( \frac{d}{dt} B))
\ + \ (A,W,B)
\ee
where A and B are the associative operators (operators of the
holonomic functions).
This lead to compatibility of quantum equations of motion for dissipative
systems and canonical commutation relations.

{\bf 4.3.} Let us define the canonical (unitary)
transformation \cite{Dir} of an operator $ \ A_H (t) = A (t) $ in the form
$ A_S (t,t_0) = S^{\dagger} (t,t_0) A_H (t) S(t,t_0) $
The operator $ A_S(t,t_0) $ satisfies the condition
$ A_S ( t_0,t_0 ) = A_H ( t_0 ) $.
In this case the equations (\ref{11}), (\ref{12}) take the form
\be
\frac{d}{dt} A_S (t,t_0) \ = \ S^{\dagger} (t,t_0)
\frac{\partial A_H (t)}{\partial t} S(t,t_0)
\ee
\be
 \label{14}
 \frac{d}{dt} \rho_S (t,t_0) = \frac{\imath}{\hbar} [ \rho_S , \ (H-W)_S ]
- \frac{1}{2} [ \Omega_S , \ \rho_S ]_{+}
\ee
This is the dissipative analogue of the Schroedinger
equations, the operators $ A_H (t) $ and $ A_S (t) $ called Heisenberg and
Schroedinger representations accordingly. The solution of the equation
(\ref{14}) is given by
\be
\rho_S (t,t_0) = U_S^{\dagger} (t,t^{\prime}) \rho_S (t^{\prime},t_0)
U_S (t,t^{\prime})
\ee
where
\be
U_S^{\dagger} (t,t^{\prime}) = Texp \frac{- \imath}{\hbar}
\int_{t^{\prime}}^{t} d \tau \ (H - W - \frac{\imath \hbar}{2}
\Omega)_S (\tau , t_0)
\ee

{\bf 4.4.}
Let us consider some important features of the basis vectors \cite{Dir}.
Account is to be taken of the time dependence of the state operator
\[ \rho_H (t) = \sum_a \rho_a [\psi_a,t>_H <\psi_a,t]_H  \]
and of the wave vectors in the Heisenberg representation $ [\psi, t>_H $. That
is  $ \ [q, t_1 >_H \ \not \equiv  \ [q, t_2 >_H $ contrary to usual
quantum mechanics. Let us define the basis vectors $ \{ [q, t> \} \  $
\cite{Dir} at the fixed time point $ t = t_f $ :
\[ 1) \quad Q_H (t)[q,t>_H = [q,t>_H q_f \qquad
2) \quad <q,t]_H [q^{\prime},t>_H = \delta (q - q^{\prime}) \]
\[ 3) \quad \int dq \ [q,t>_H <q,t]_H = 1 \qquad
4) \quad Q_H (t) = \int dq \ [q,t>_H  q_f <q,t]_H \]
\[ 5) \quad [\psi, t>_H = \int dq \ [q,t_f>_H \Psi_H (q,t,t_f) \]
where $ \Psi_H (q,t,t_f) = \ <q,t_f]_H [\psi ,t >_H $. It is easy to
prove the following statements:  1. The basis vector unitary
transformed is a basis vector ; 2. There exists a unitary
transformation for any two basis vectors defined at the non equal
time points.  Thus, Schroedinger representation of
the basis vector $ [q , t, t_0 >_S \equiv S^{\dagger} (t - t_0) [q ,t >_H $
might be considered as the unitary transformation of the basis vector
\[ [ q , t_0 >_H = S^{\dagger}  (t - t_0 ) [ q , t >_H = [ q , t , t_0 >_S \]

therefore the trace of the operator can be defined only in fixed
time point. Note that the operator of the state $ \rho (t) $
satisfies the usual condition
\[ Sp_t ( \ \rho (t) \ ) \ = \ 1 \ \ \ \ ( \forall t=t_{fixed}) \]
 where we take into account the fixed-time definition of
the basic vectors. The average of the physical quantity
$A(t)=A(p,q,t)$ is defined by
\be
 < A(t) >_t \ = \ Sp_t ( \ A(t) \rho(t) \ ) \ \ \ ( \forall t=t_{fixed})
\ee
and the time derivative of the average quantity can be defined only
by following
\be
\frac{d}{dt} <A(t) >_{\tau} \  \equiv \ \frac{d}{dt}
Sp_{\tau} ( \rho( \tau) A(t)) \ \ \ (\forall t= t_{fixed})
\ee
i.e. as the average of the time derivative  of the operator.

{\bf 4.5.} Let us consider now Green's functions and its Feynman
representation. If we take into account the equation (\ref{14}) we can
write the dissipative Schroedinger equation for wave vector in the form
\be
\imath \hbar \frac{d}{dt} [ \psi ,t,t_0 >_S = ( H - W -
\frac{\imath \hbar}{2} \Omega )_S (t, t_0) [\psi , t , t_0 >_S
\ee
The simple example of this equation for the harmonic oscillator with friction
is considered in  {\bf 4.7.} .
Account is to be taken of the time dependence of state in Heisenberg
representation. Therefore we make the distinctions between following Green's
functions
\be
\Psi_S (q,t) = \int dq^{\prime} \ G_S (q,q^{\prime},t - t^{\prime})
\ \Psi_S (q^{\prime}, t^{\prime})
\ee
\be
\Psi_H (q,t) = \int dq^{\prime} \ G_H (q,q^{\prime},t - t^{\prime})
\ \Psi_H (q^{\prime},t^{\prime})
\ee
where
\be
 \Psi_S (q,t)  \ \equiv \ <q,t]_S [\psi , t>_S \ \equiv \
<q,t]_H [\psi,t>_H
\ee
\be
\Psi_H (q,t) \ \equiv \ <q,t]_H [\psi , t>_S \ \equiv \
<q,t]_S [\psi ,t>_H
\ee
\[ G_S (q,q^{\prime},t-t^{\prime}) \equiv \ <q,t]_S \
U_S^{\dagger} (t,t^{\prime}) \
[q^{\prime}, t^{\prime}>_S \theta (t-t^{\prime}) \ \equiv \]
\be
\equiv \ <q,t]_H \ U_H^{\dagger} (t,t^{\prime}) \
[q^{\prime},t^{\prime}>_H \theta (t-t^{\prime})
\ee
\be
G_H (q, q^{\prime},t-t^{\prime}) \equiv \ <q,t]_S \
U_H^{\dagger} (t-t^{\prime})
 \ [q^{\prime} , t^{\prime} >_S \theta(t-t^{\prime})
\ee
and $ \  [q,t>_H \ \equiv \ [q,t=t_{fixed} >_H \ . $
The Green's function satisfies the time-dependent equation
\be
\imath \frac{d}{dt} G_S (q,q^{\prime},t) = ( H - W -
\frac{\imath \hbar}{2} \Omega )_S
G_S (q,q^{\prime},t)  \ \ and  \  G_S (q,q^{\prime}, 0) =
\delta (q-q^{\prime})
\ee
If we use the method considered in \cite{Fad} and the conditions
\be
<p,t_f]_H ( H - W - \frac{\imath \hbar}{2} \Omega )_H [q,t_f>_H =
( h - w - \frac{\imath \hbar}{2}
\omega) (q_f , p_f ) <p,t_f]_H [q,t_f>_H
\ee
\be
<q^{n+1},t_n]_S U_S^{\dagger} (t_{n+1} - t_n ) [q^n,t_n >_S \simeq
\ee
\be
\simeq <q^{n+1}, t_n]_H exp \frac{- \imath (t_{n+1} - t_n)}{\hbar}
(H - W - \frac{\imath \hbar}{2} \Omega )_H (t_n) [q^n,t_n >_H \ ,
\ee
then the Feynman representation  of the Green's functions is given by
\be
\label{15}
G_S (q,q^{\prime},t - t^{\prime}) = \int Dq \ Dp \ exp \frac{\imath}{\hbar}
\int_{t^{\prime}}^{t} d \tau \
( p \frac{dq}{d \tau } - h (q,p,\tau ) + w (q,p, \tau)
+ \frac{\imath \hbar }{2} \omega  )
\ee
In the same way we can formulate the path integration and generating
functional in the quantum field theory  which was
considered in the papers \cite{Tar2,Tarpl,Tarmpl}.
As one of methods to solve the
quantum dissipative equations we suggested the normal geodesic
coordinate and covariant background field method
\cite{Veb2,Witt,Mukh}
which is generalization of the Teylor's series
expansion around the classical solution and was considered for
Riemannian \cite{Witt,Mukh}, affine
\cite{Veb2,Veb1} and
affine-metric \cite{Tarb1,Tarpl,Tarmpl,Tartmf2} manifolds.

{\bf 4.6.} Let us consider the harmonic oscillator with  friction
and prove that the configurational space of this oscillator is curved
by the quantum fluctuations.
It is known that the harmonic oscillator configurational space metric
is defined by the kinetic energy \cite{Lanc}
\be
d^2 s = 2T (dt)^2 = \delta_{ij} dq^i dq^j
\ee
where $ T =  \sum \nolimits_{i}^{n} (1 / 2) (dq^i/dt)^2 $ is the
kinetic energy of the harmonic oscillator.
The existence of the potential forces with the potential
\[ U(q) =  \sum \nolimits_{i=1}^{n}
( \omega^2_i / 2) (q^i)^2 ; \ \ \omega^2_i = \sqrt{k_i/m_i} \]
lead to the deformation of the configurational space metric
in the form
\[ d^2 s =(E - U(q)) \delta_{ij} dq^i dq^j \]
for conservative systems only \cite{Lanc}, where E is the total energy.
The dissipative forces don't allow us to derive the mechanical
trajectory by Jacobi variational principle \cite{Lanc}. That is the
configuration space of the harmonic oscillator with friction is flat.
The equations of motion for this oscillator in
n-dimensional configuration space are
\be
\label{16}
 du^i/dt + \omega^2_i q^i = f_i (q,u)
\ee
where $ i;j = 1,...,n ; u^i \equiv { {dq^i}/{dt}} $ and $ f_i(q,u)$
is the dissipative force. Let us consider the following friction
force
\be
\label{17}
f_i(q,u) = c_{ij} (q) u^j + D_{ikl}(q) u^k u^l
\ee
As is well known, the equation (\ref{16}) with friction (\ref{17}) cannot be
derived from the least action principle.
The nonholonomic functional for this harmonic oscillator with friction is
defined by the nonholonomic equation linear in the variation $ \delta q^k $
\be
\delta \tilde W = \delta \int dt \ w(q,u) = \int dt \ f_k (q,u) \delta q^k
\ee

We derive the background field expansion of the Hamiltonian h(q,p),
nonholonomic functional $ w $, and omega $ \omega (q,p) $
around the classical solution $ q^i_0 $  of this oscillator
as a power series in $ \xi^i(t) : q^i(t) = q^i_0 (t) + \xi^i (t) $.
The functional integral (\ref{15}) over momentum $p^i$ is Gaussian integral
and we derive the path integral form of Green's function
\be
 G_S (q,q^{\prime},t - t^{\prime}) = N \int D \xi^i \ exp
\frac{\imath}{\hbar} \int_{t^{\prime}}^{t} d \tau \
( T (q_0) + Z_1 (q_0,\xi,\tau ) + Z_2 (q_0,\xi, \tau)  )
\ee
where $ \ T(q_0) = \frac{1}{2} \delta_{ij} v^i_0 v^j_0 \ ; \ v^i_0 =
d q^i_0 / d \tau $;
\be
Z_1(q_0,\xi) = \frac{1}{2} \delta_{ij} \frac{d \xi^i}{d \tau}
\frac{d \xi^j}{d \tau} \ - 2D_{ikj}(q_0) v^k_0 \xi^i \frac{d \xi^j}{d
\tau} - D_{inm;j}(q_0) v^n_0 v^m_0 \xi^i \xi^j - \frac{\omega^2_i}{2}
(\xi^i)^2
\ee
and $ a_{n;m} (q) \equiv \partial a_n(q) / \partial q_m \ $. $ \ Z_2
(q_0,\xi) \ $ is the sum of the series terms in $ \xi $ which don't
contribute to the one-loop finite renormalization of the metric.
In the calculation of
the vacuum diagrams $q^i_0 (t) $ is regarded as an external field and
$ \xi^i (t) $ is the quantum field. The vacuum contributions to the
Green's function which have the form $ (1/2) T_{ij} (q_0) v^i_0 v^j_0 $
lead to the metric redefinition
\[ d^2 s = 2T (q_0)(dt)^2 = ( \delta_{ij} + T_{ij}) dq^i_0 dq^j_0 \]
The one-loop vacuum contribution for the harmonic oscillator with
friction (\ref{17}) has the form
\be
T_{ij} = \sum_{k=1}^{n} \ ( \ \sum_{l=1}^{n} \frac{2}{ \omega_k +
\omega_l }  D_{kli} (D_{klj} - D_{lkj}) - \frac{1}{\omega_k} D_{ikk;j} )
\ee
It is easy to see that the background configurational space of the
quantum harmonic
oscillator with friction (\ref{17}) is not flat. If the friction is
quadratic in the velocity with coefficients which depend on the
coordinates or these coefficients are non completely symmetric tensor the
configurational space is curved by the quantum fluctuations. The full
expression of the two-loop metric redefinition terms is complicated.
Let us note the simple condition which allow the configurational space
to be flat, no matter one-loop or two-loop vacuum diagrams are
considered, is given by
\be
D_{ikl;j} = 0 \quad and \quad D_{ikl} = D_{(ikl)}
\ee

{\bf 4.7.} In the conclusion, let us consider the dissipative analogue
of Schrodinger equation for one dimensional harmonic
oscillator with friction. Hamiltonian and nonholonomic functional are
following
\[ h= \frac{p^2}{2m} + \frac{m \omega^2 q^2}{2} \ ; \quad \delta w =
\gamma m p \ \delta q \]
We use the background field method and expand the Hamiltonian and
nonholonomic functional in Taylor series of $ Q = q - q_0 $,
where $ q_0 $ is the
solution of the classical equation of motion in the coordinate space.
Let us choose $ q_0 = 0 $. In this case the generalized
Schroedinger equation takes the form
\be
 \imath \hbar \frac{d}{dt} \Psi (t) = [-\frac{\hbar}{2m}
\frac{\partial^2}{\partial Q^2} +
\ \imath \hbar \gamma \ Q
\frac{\partial}{\partial Q } \ + \ \frac{m \omega^2}{2} Q^2 -
\frac{\imath }{2} \gamma \ ]  \ \Psi (t)
\ee
The stationary state $ \Psi (\xi, t) = u (\xi)
exp - \frac{\imath}{\hbar} E t $ is defined by the equation
\[  u^{\prime \prime} (\xi) - a \xi u^{\prime} (\xi)  + (\varepsilon
- \xi^2 ) u(\xi) = 0 \]
where
\[ a= \frac{2 \imath \gamma }{\omega} \ ; \ \ \xi =
\sqrt{\frac{m\omega}{\hbar}} Q \ ; \ \ \varepsilon = \frac{2}{\hbar \omega}
(E -\frac{\imath}{2} \gamma) \]
Let us consider the function $u(\xi)$ in the form
\[ u(\xi) = (\sum_{k=0}^{n} A_k \xi^k) \ exp -\frac{1}{2} s \xi^2 \]
where s is the solution of the equation $ s^2 + a s - 1 = 0 $ and
$ n < \infty $ .

As a result we obtain the following eigenvalues
\[ E_n = \hbar \sqrt{\omega^2 - \gamma^2} (n + \frac{1}{2}) - \imath \gamma \]
when $ 0 < {\gamma^2}/{\omega^2} < \frac{1}{2} $ and the continuous spectrum,
when $ {\gamma^2}/{\omega^2} > \frac{1}{2} $. Note that the life time for
the state is $ T = \hbar / 2 \gamma  <  \infty $. We can rewrite the result
in the form
\[ \Delta E_n (\omega) = ( \hbar \sqrt{\omega^2 - \gamma^2} \ when \
\omega^2 > 2 \gamma^2 ) \bigwedge (0 \ when \ \omega^2 < 2 \gamma^2 ) \]
where $\Delta E_n  \equiv E_n - E_{n-1}$.
Note that the jump in the point $ \omega_0 = \sqrt{2} \gamma $ is the purely
quantum dissipative effect.

\vskip 1 cm

\centerline{* \ \ \ \ \ * \ \ \ \ \ *}

The author would like to thank Professor Abdus Salam, the
International Atomic Energy Agency and UNESCO for hospitality
at the International Centre for Theoretical Physics, Trieste.

This work was partially supported by Russian Fund for Fundamental Researches
grant N. 94-02-05869.

\end{document}